\documentclass[12pt,a4paper]{article}
\usepackage{amssymb,amsthm}
\newcommand{\de}{\mathrm{d}}
\newcommand{\ii}{\mathrm{i}}
\newcommand{\cO}{{\mathcal{O}}}
\newcommand{\cC}{{\mathcal{C}}}
\newcommand{\PP}{{\mathbb{P}}}
\newtheorem{thm}{Theorem}
\newtheorem{lemma}[thm]{Lemma}
\newtheorem{prop}[thm]{Proposition}

\title{Normal bundles to Laufer rational curves \\ in local Calabi-Yau threefolds}
\author{U. Bruzzo and A. Ricco\\[-5pt] \footnotesize  International School for Advanced Studies (SISSA/ISAS), \\[-7pt]   \footnotesize Via Beirut 2, 34014 Trieste, Italy\\[-7pt]
\footnotesize  Istituto Nazionale di Fisica Nucleare, Sezione di Trieste}

\date{\small 16 November 2005}

\begin{document}

\maketitle 

\thispagestyle{empty}

\begin{abstract}   
We prove a conjecture by F.~Ferrari. Let $X$ be the total space of a nonlinear   deformation
 of a rank 2 holomorphic vector bundle on a smooth rational curve,
such that $X$   has trivial canonical bundle and has sections. Then
the normal bundle to such sections is computed in terms
of the rank of the Hessian of a suitably defined superpotential  at its
critical points.
\end{abstract}




\paragraph{Introduction.}
In this paper we consider particular embeddings of smooth rational curves in local Calabi-Yau threefolds, called Laufer curves \cite{laufer, Katz:2000ab}. 
 (We use the definition of the physics community, calling Calabi-Yau a quasi-projective threefold with trivial canonical bundle; the term ``local'' refers to non-compactness.)
  These geometries have shown to be very useful to understand several features of string theories and supersymmetric gauge theories.
In particular they are relevant for brane dynamics and geometric transition/large $N$ dualities. Geometric transition interprets the resummation of the open string sector of an open-closed string theory as a transition in the target space geometry, connecting two different components of a moduli space of Calabi-Yau threefolds.
The local Calabi-Yau that we consider represents the open string side of conjectured geometric transitions. In particular, open topological B-type strings in these geometry reduces to matrix models in which the parameters of the complex structure are the coupling constants. 

This is directly connected (via F-terms) to the possibility of geometrically engineering supersymmetric gauge theories in Type IIB string theory. 
Let $\mathbb{R}^4 \times X$ be the target space of the theory, where $X$ is a Calabi-Yau threefold, and $\mathcal{C}$ a rational curve in $X$, with normal bundle $V$ and $N$ D5 branes wrapped on it. The effective field theory is a  $\mathcal{N}=1$ supersymmetric gauge theory with gauge group $U(N)$.
The space of vacua of this gauge theory, given by the critical points of the effective superpotential, is locally described by the versal deformation space of the curve in $X$. For a given vacua, there are $h^0$ massless chiral superfields in the adjoint representation of the gauge group, where $h^0 := \dim H^0(\mathcal{C}, V)$. On the other hand, the number of massless chiral multiplets is equal to the corank of the Hessian of the superpotential at this vacuum. This relation led \cite{Ferrari:2003vp} to conjecture the result expressed in our Proposition \ref{prop:ferrari}.

For an account of these aspects, see also \cite{Curto, Mazzucato:2005fe, Bonelli:2005dc} and references therein.

From a strictly mathematical viewpoint, the problem is the following.
Let $V$ be a rank-2 holomorphic vector bundle on a rational curve $\cC$
such that its total space has trivial canonical bundle,
and assume that $V$ has a global section. We deform $V$ to a nonlinear
fibration $X$ in such a way that $X$ still has trivial canonical bundle
and the fibration has sections. The normal bundle to such a section
of course splits as a direct sum of two line bundles in view of Grothendieck's
classification of vector bundles on curves of genus zero \cite{Gro}. The problem
is to compute these line bundles. The solution is obtained in terms
of a superpotential $W$ than one associates with the deformations of $V$:
the sections of $X$ are given by the critical points of $W$, and 
the degrees of the above mentioned  line bundles are given, in accordance
with a conjecture by Ferrari \cite{Ferrari:2003vp}, by the rank of the
Hessian of $W$ at those critical points.

\paragraph{Definition of $X$.}
Let $\cC \simeq\PP^1$ be a smooth rational curve and $V \to \cC$ a rank-2 holomorphic vector bundle on $\cC$, with $\det V \simeq K_{\cC} \simeq \cO(-2)$, so that 
 the total space of the bundle $V$  has trivial canonical bundle. 
Then $V \simeq  \cO(-n-2) \oplus \cO(n)$ for some $n$. 
We  consider deformations of $V$ given in terms of transition functions
in the standard atlas $\mathcal{U} = \{ U_0, U_1 \}$ of $\mathbb{P}^1$ 
 as 
\begin{equation}\label{eq:fibration}
\left\{\begin{array}{rcl}
   z'  &=& 1/z  \\[3pt]
   \omega_1' &=& z^{-n}  \omega_1   \\[3pt]
   \omega_2' &=& z^{n+2} \left( \omega_2 + \partial_{\omega_1} B(z, \omega_1) \right)\ .
\end{array}\right.\end{equation}
Note that the complex manifold $X$ defined as the total space of this fibration has again trivial canonical bundle.  The term
$B(z, \omega)$ is a holomorphic function on $(U_0 \cap U_1) \times \mathbb{C}$ and is called the \emph{geometric potential}. If we expand the function $B$ in its second variable
\begin{equation}
B(z,\omega_1)=\sum_{d=1}^\infty \sigma_d(z)\,\omega_1^d
\end{equation}
each coefficient $\sigma_d$ may be regarded as a cocycle
defining an element in the group
\begin{equation}
H^1(\PP^1,\cO(-2-dn))\simeq H^0(\PP^1,\cO(nd))^\ast\,.
\end{equation} 

\paragraph{The  superpotential.} If we consider $\cC$ as embedded in $X$ as its zero section, and consider the problem of deforming the pair $(X\,\cC)$, the space
of versal deformations can be conveniently described by a superpotential \cite{Katz:2000ab}. In the case at hand the  superpotential $W$ can be defined as
the function of $n+1$ complex variables given by
\begin{equation}\label{eq:superpotential}     W(x_0, \dots, x_n) = \frac{1}{2 \pi \mathrm{i}}
 \oint_{\cC_0} B\left(z,\omega_1(z)  \right) \,\de z
\end{equation}
where $z$ and $z'$ are local coordinates on $U_0$ and $U_1$,
and the parameters $x_0, \dots, x_n$ define sections of 
 the   line bundle $\cO(n)$ by letting \begin{eqnarray}\label{eq:section1}
   \omega_1(z)= \sum_{i=0}^{n} x_i z^i \ , \qquad \omega_1'(z') = \sum_{i=0}^{n} x_i (z')^{n-i}\,.
\end{eqnarray} One should note that the superpotential $W$ can be obtained by applying
to the function $B$, regarded as an element in $H^0(\PP^1,\cO(nd))^\ast$, the dual of the
multiplication morphism
\begin{equation}
H^0(\PP^1,\cO(n))^{\otimes d} \to H^0(\PP^1,\cO(nd))
\end{equation}
(here one should regard the dual of $H^0(\PP^1,\cO(nd))$ as a space
of Laurent tails).

The key to the result  we want to prove is the relationship occuring between
the superpotential $W$ and the sections of the fibration $X\to \cC$
(cf.~\cite{Katz:2000ab, Ferrari:2003vp}).

\begin{lemma}\label{lemma:sect}
The holomorphic sections of the   fibration $X\to \cC$ 
are in a one-to-one correspondence with the critical points of the superpotential, \emph{i.e.}, with the  solutions of the equations
\begin{eqnarray}
\frac{\partial W}{\partial x_i} = 0 \ , \quad i=0, \dots, n \ .
\end{eqnarray} 
\end{lemma}
\begin{proof}
This  can be verified by explicit calculations \cite{Ferrari:2003vp} after representing  the sections    of $X$ as
\begin{equation}\label{eq:section2}\begin{array}{rcl} 
 \omega_2(z) &=& \displaystyle  -  \frac{1}{2 \ii \pi}\oint_{C_z} \frac{\partial_\omega B (u,\omega_1(u))}{u-z}  \,  \de u \\[12pt]
 \omega_2'(z') &=& \displaystyle  \frac{1}{2 \ii \pi} \oint_{C_{z'}}
 \frac{\partial_\omega B (1/u,\omega_1(1/u))}{u^{n+2}(u-z)}  \,\de u
\end{array}\end{equation}
where the contour $C_z$ (resp. $C_{z'}$)  encircles the points $0$ and $z$ (resp $z'$). So (\ref{eq:section1}) and (\ref{eq:section2}) yield a rational curve $\Sigma \subset X$ for each critical point $(x_0, \dots, x_n)$ of $W$.
\end{proof}

\paragraph{Ferrari's Conjecture.}
Now we state and prove Ferrari's conjecture.
\begin{prop} \label{prop:ferrari}
The normal bundle to the section $\Sigma$ of $X$ determined by a critical point $(x_0, \dots, x_n)$ of $W$ is $\cO_\Sigma(-r-1) \oplus \cO_\Sigma(r-1)$ where $r$ is the corank of the Hessian of $W$ at that point.
\end{prop}

To calculate the normal bundle to $\Sigma$ we first need to linearize the transition functions around the given section.
Defining new coordinates $\delta_i = \omega_i - \omega_i(z)$, $\delta_i' = \omega_i' - \omega_i'(z)$, we obtain
\begin{eqnarray}
  \delta_2' =  z^{n+2} \left(\delta_2 +  h(z) \delta_1 + g(z) \right)
\end{eqnarray}
where 
\begin{eqnarray} 
   g(z) = \partial_\omega B(z, \omega_1(z)) \ ,  \qquad
   h(z) =  \partial^2_\omega B(z, \omega_1(z)) 
\end{eqnarray}
and at a critical point of $W$ we have $g(z) = 0$ using relation (\ref{eq:derivatives}) in the appendix. Furthermore, again from (\ref{eq:derivatives}), for $h(z)$ we have
\begin{eqnarray}\label{eq:cocycleW}
  h(z) = \sum_{i \leq j =0}^{n} \partial_i \partial_j W^{(k)}_{d} z^{-(i+j)-1}
\end{eqnarray}
up to terms that can be can be readsorbed by holomorphic change of coordinates (see the Appendix).

Now we need the following. Let us consider an extension of vector bundles on $\PP^1$ of the form
\begin{eqnarray}
0 \longrightarrow \cO_{\PP^1}(-n-2) \longrightarrow \Phi 
  \longrightarrow \cO_{\PP^1}(n) \longrightarrow 0
\end{eqnarray}
parametrized by a cocycle $\sigma \in H^1 (\PP^1, \cO_{\PP^1}(-2n-2))$. With respect to the two standard charts $U_0, U_1$ and in the coordinate $z$ of $U_0$,  $\sigma$ can be written as
\begin{eqnarray}
   \sigma(z) = \sum_{k=0}^{2n} \widetilde{t}_k z^{-k-1} \ .
\end{eqnarray}
Let us define a quadratic form (quadratic superpotential)
on the global sections   of  the line bundle $\cO_{\PP^1}(n)$:
\begin{eqnarray}
   H(x_0, \dots, x_n) = \sum_{k=0}^{2n} \widetilde{t}_k \sum_{i,j=0 \atop i+j=k }^{n} x_i x_{j} = \sum_{i,j=0}^{n} H_{ij} x_i x_j \ .
\end{eqnarray}

\begin{lemma}
The vector bundle $\Phi$ is $\cO_{\mathbb{P}^1}(r-1) \oplus \cO_{\PP^1}(-r-1)$, where $r$ is the corank of the quadratic form $H$.
\end{lemma}

\begin{proof} By Lemma \ref{lemma:sect} the sections of the bundle $\Phi$ correspond to the critical points of $H$, \emph{i.e.,} to the solutions of the linear system
\begin{eqnarray}
   \sum_{j=0}^{n} H_{ij} x_j = 0 \ .
\end{eqnarray}
The dimension of this space is $r$, the corank of $H$. The only rank two vector bundle over $\mathbb{P}^1$ with determinant $\cO_{\PP^1}(-2)$ and $r$ linearly indipendent holomorphic sections is $\cO_{\PP^1}(r-1) \oplus \cO_{\mathbb{P}^1}(-r-1)$.  \end{proof}

The proof of   Proposition \ref{prop:ferrari} is now complete: 
in fact, by (\ref{eq:cocycleW}) the quadratic form $H$ corresponds to the Hessian of the superpotential $W$ at its critical points.

\appendix
\section{Some formulas for the potentials}
We group here some formulas that turn out to be useful in checking the
computations involved in the results presented in this paper.
\paragraph{The geometric potential.}
The geometric potential (deformation term) $B(z, \omega_1)$ is holomorphic on $\mathbb{C}^* \times \mathbb{C}$ and can be cast in the form
\begin{eqnarray}\label{eq:expansionB}
 B(z, \omega) =  \sum_{d=0}^{\infty} \sum_{k=0}^{dn} t^{(k)}_d  B^{(k)}_d (z, \omega)
\end{eqnarray}
where 
\begin{eqnarray}
 B^{(k)}_d (z, \omega) = z^{-k-1} \omega^d \qquad k= 0, \dots, dn \ .
\end{eqnarray}
The terms with $k < 0$ or $k >  dn$ can be readsorbed by a holomorphic change of coordinates. For $l := - k -1 \geq 0$, we define $\widetilde{\omega}_2 := \omega_2 + d z^{l} \omega_1^{d-1}$, and for $m := k - dn -1 \geq 0$, we define
\begin{eqnarray}
\widetilde{\omega}_2':= \omega_2' - (z')^m (\omega_1')^{d-1}\ .
\end{eqnarray}

\paragraph{The superpotential}
The superpotential that corresponds to $B^{(k)}_d$, given by (\ref{eq:superpotential}), is
\begin{eqnarray}\label{eq:expansionW}
 W^{(k)}_{d}(x_0, \dots, x_n) = \sum_{i_1, \dots, i_d=0 \atop i_1+ \dots + i_d=k }^n 
                                x_{i_1} \dots x_{i_d} \ .
\end{eqnarray}
We can obtain simple relations for the derivatives of these polynomials:
\begin{eqnarray}
 \frac{\partial W^{(k)}_{d}}{\partial x_j}  &=& \sum_{i_1, \dots, i_d=0 \atop i_1+ \dots +  i_d=k }^n 
                                                d \left( \frac{\partial x_{i_1}}{\partial x_j} x_{i_2} \dots x_{i_d} \right) \nonumber \\
                                            &=& d \sum_{i_1, \dots, i_{d-1}=0 \atop i_1+ \dots +  i_{d-1}=k-j }^n x_{i_1} 
                                                \dots x_{i_{d-1}} 
                                            \ = \ d W^{(k-j)}_{d-1}
\end{eqnarray}
and in general we have
\begin{eqnarray}
 \frac{\partial}{\partial x_{j_1}}\dots \frac{\partial}{\partial x_{j_l}} W^{(k)}_{d} = d (d-1)\dots (d-l+1) W^{(k -j_1\dots- j_l)}_{d-l}
\end{eqnarray}

\paragraph{Relations between the derivatives of the potentials}
Given a section $\omega_1(z)$, we have
\begin{eqnarray}\label{eq:derivatives}
\partial_\omega B(z, \omega_1(z))  = \sum_{j=0}^n \frac{\partial W}{\partial x_j} z^{-j-1} + \mathrm{trivial\ terms} \nonumber \\
\partial^2_\omega B(z, \omega_1(z)) = \sum_{i \leq j =0}^{n} \partial_i \partial_j W z^{-(i+j)-1} + \mathrm{trivial\ terms}
\end{eqnarray}
where the ``trivial terms'' can be readsorbed by a holomorphic change of coordinates. We can obtain these results from (\ref{eq:expansionB}) and (\ref{eq:expansionW}). We have
\begin{eqnarray}
\partial_\omega B^{(k)}_d (z, \omega_1(z)) = 
       d \sum _{i_1, \dots, i_{d-1} = 0}^{n} x_{i_1} \dots  x_{i_{d-1}} z^{i_1 + \dots + i_{d-1}-k-1}
\end{eqnarray}
and the only non-trivial terms are such that $0 \leq - (i_1 + \dots + i_{d-1}-k) \leq n$.
In the same way, for the second derivatives we have
\begin{eqnarray}
     \partial^2_\omega B^{(k)}_{d}(z, \omega_1(z)) = 
           d (d-1) \sum _{i_1, \dots, i_{d-2} = 0}^{n} x_{i_1} \dots  x_{i_{d-2}} z^{i_1 + \dots + i_{d-2}-k-1}
\end{eqnarray}
The relevant terms are those with  $0 \leq - (i_1 + \dots + i_{d-1}-k) \leq 2 n$.

\end{document}